\documentclass[aps, prb, superscriptaddress, reprint, nobibnotes]{revtex4-2}
\usepackage{amsmath,amssymb}
\usepackage{color}
\usepackage{comment}
\usepackage{bm,graphicx,hyperref}
\hypersetup{%
  breaklinks = {true},
  citecolor = {blue},
  colorlinks = {true},
  linkcolor = {red},}

\usepackage{lineno}

\begin{document}

\title{Three-dimensional harmonic oscillator as a quantum Otto engine}

\author{A. Rodin}
\affiliation{Yale-NUS College, 16 College Avenue West, 138527, Singapore}
\affiliation{Centre for Advanced 2D Materials, National University of Singapore, 117546, Singapore}
\affiliation{Department of Materials Science and Engineering, National University of Singapore, 117575, Singapore}

\begin{abstract}

A quantum Otto engine based on a three-dimensional harmonic oscillator is proposed.
One of the modes of this oscillator functions as the working fluid, while the other two play the role of baths.
The coupling between the working fluid and the baths is controlled using an external central potential.
All four strokes of the engine are simulated numerically, exploring the nonadiabatic effects in the compression and expansion phases, as well as the energy transfer during the working fluid's contact with the baths.
The efficiency and power of several realizations of the proposed engine are also computed with the former agreeing well with the theoretical predictions for the quantum Otto cycle.

\end{abstract}	
\maketitle

\section{Introduction}
\label{sec:Introduction}

The purpose of a quantum engine is the same as that of a classical one: converting energy into work.
The main distinction is that quantum components play the role of the working fluid.~\cite{Bhattacharjee2021, Myers2022}
Examples of quantum systems that can act as the fluid include two-level systems~\citep{Huang2012, Thomas2018}, one~\citep{DelCampo2014, Rossnagel2014, Kosloff2017, Cavaliere2022, Fei2022} or multiple~\citep{Boubakour2023} harmonic oscillators, or photons~\citep{Zhang2014}.
The second difference between classical and quantum varieties is their energy sources.
While classical versions get their energy from hot reservoirs, quantum engines can be powered by measurements, as described in several recent studies.~\citep{Chand2017, Elouard2017, Elouard2017a, Chand2018, Elouard2018, Das2019, Bresque2021, Manikandan2022, Alam2022, Jussiau2023}

Despite the variety of quantum engines, perhaps the most common one is based on the classical Otto cycle, where the working fluid is a gas.
This cycle involves four processes: adiabatic compression, constant-volume heat transfer to the gas, adiabatic expansion, and constant-volume heat transfer away from the gas.
The two heat exchange processes indicate that the engine requires two heat reservoirs for its operation.
In the quantum variety of the Otto cycle, the working fluid is often chosen to be a harmonic oscillator~\citep{DelCampo2014, Rossnagel2014, Kosloff2017, Cavaliere2022, Fei2022, Boubakour2023} because the compression and expansion can be achieved by adjusting the oscillator's force constant.

This work proposes an implementation of a quantum Otto engine, consisting of a three-dimensional harmonic oscillator with anisotropic force constants in the three orthogonal directions.
The three corresponding modes act as a hot reservoir, a cold reservoir, and the ``gas" with an adjustable force constant.
The coupling between the gas and the baths can be controlled using an external central potential, as shown in Ref.~\citep{Rodin2023}.
Although single ``bath" modes do not function as true thermodynamic reservoirs, they can be regarded as ancillae that deliver energy to or extract it from the gas mode.
While a particular bath mode is disconnected from the gas, it can be either cooled or heated to return it to its pre-contact state.
The cooling and heating mechanisms for the ancillae may involve optics (such as laser cooling) or coupling the mode to a thermodynamic reservoir at a particular temperature.
For the sake of brevity, the two ancillary modes will be referred to as ``baths" in the text while keeping in mind that there are additional external reservoirs.

A reasonable question is why the auxiliary bath modes are necessary instead of directly cooling and heating the gas mode using the same means.
The reason is complexity: cooling and heating of a single mode means one must be able to swap the heating and the cooling mechanisms.
If there are dedicated ancillary modes, however, each one of them needs to be connected to, at most, a single external system.
Moreover, if the ``resetting" process of the bath modes is sufficiently slower than their rate of energy exchange with the gas mode, the external system can remain connected to the bath mode throughout the engine's operation.
The idea of always keeping the bath connection on for different implementations of the quantum Otto cycle was explored earlier in Refs.~\citep{Chand2017a, Chand2021}.

The efficiency of an ideal quantum Otto cycle is obtained by assuming that the working fluid is compressed and expanded adiabatically, and that it reaches the bath temperature.~\citep{Bhattacharjee2021}
For an engine to deliver a nonvanishing power, they must be able to complete their cycles in a finite amount of time.
Therefore, the compression and expansion phases are likely to give rise to nonadiabatic effects.
Additionally, for the engine proposed here, one does not expect the gas mode to reach the temperature of the bath mode.
In fact, the state of the gas mode is likely not to be thermal generally.
Nevertheless, this work demonstrates that, despite these deviations from the ideal cycle, the resultant efficiency can be close to the predicted value.

To show that the engine proposed does not need to adhere to the ideal configuration to deliver such efficiency, it is useful to first focus on individual strokes of the Otto cycle.
Thus, after presenting the theoretical model describing the engine in Sec.~\ref{sec:Model}, this work dedicates Secs.~\ref{sec:Compression_and_expansion} and \ref{sec:Bath_coupling} to the compression/expansion phases and energy exchange with the baths, respectively.
The entire cycle of the engine operation is presented in Sec.~\ref{sec:Engine_operation}.
Summary and conclusions are found in Sec.~\ref{sec:Summary}.

All computations are performed using the {\scshape julia} programming language.~\citep{Bezanson2017}
The plots are made using Makie.jl package~\citep{Danisch2021} using the color scheme designed for colorblind readers.~\citep{Wong2011}
The scripts used for computing and plotting can be found at https://github.com/rodin-physics/quantum-oscillator-engine.

\section{Model}
\label{sec:Model}

As discussed in the introduction, the engine consists of a three-dimensional harmonic oscillator, where two of the dimensions function as baths, and the remaining one operates as the ``gas" in the Otto cycle.
The Hamiltonian for such a system is given by

\begin{align}
    \hat{H}(t) &= \sum_{d = c,h,g}\left(\frac{ \hat{p}_d^2}{2m}+\frac{k_d}{2}\hat{x}_d^2\right)
    +
    \frac{\kappa(t)}{2}\hat{x}^2_g
    \nonumber
    \\
    & +U_h(\hat{x}_h,\hat{x}_g, t)
    + U_c(\hat{x}_c,\hat{x}_g, t)\,,
    \label{eqn:Hamiltonian}
\end{align}
where the $c$, $h$, and $g$ correspond to \textbf{c}old, \textbf{h}ot, and \textbf{g}as, respectively.
The last term of the first line allows the gas to be compressed and expanded, as required by the engine operation.
The second line contains the interaction terms between the gas and the baths.

The most natural way to describe the system is using the Fock basis $|j\rangle_c\otimes|k\rangle_h\otimes |l\rangle_g\equiv |j,k,l\rangle$, where $j$, $k$, and $l$ are the energy levels of the three modes.
Therefore, one may be tempted to write the portions of Eq.~\eqref{eqn:Hamiltonian} corresponding to the baths as $\hbar\Omega_b(\hat{b}^\dagger b + 1/2)$.
Although entirely valid, this choice makes the subsequent computations messier.
Consequently, it is easier first to rewrite Eq.~\eqref {eqn:Hamiltonian} as

\begin{align}
    \hat{H}(t) &= \sum_{d = c,h,g}\left(\frac{ \hat{p}_d^2}{2m}+\frac{k}{2}\hat{x}_d^2\right)
    +
    \frac{\kappa(t)}{2}\hat{x}^2_g
    +
    \frac{\kappa_\mathrm{max}}{2}\hat{x}^2_h
    \nonumber
    \\
    & +U_h(\hat{x}_h,\hat{x}_g, t)
    + U_c(\hat{x}_c,\hat{x}_g, t)\,.
    \label{eqn:Hamiltonian_Rewrite}
\end{align}
Here, it was assumed that the gas force constant takes a range of values $k_c = k \leq k +\kappa(t)\leq k_h = k + \kappa_\mathrm{max}$ during the compression and expansion phases.
Writing the first line of Eq.~\eqref{eqn:Hamiltonian_Rewrite} using the ladder operators gives the following Hamiltonian

\begin{align}
    \hat{H}(t) &= \sum_{d = c,h,g}\hbar \Omega\left(\hat{d}^\dagger \hat{d} + \frac{1}{2}\right)
    \nonumber
    \\
    &+
    \alpha(t)\frac{\hbar\Omega}{4}(\hat{g}^\dagger+\hat{g})^2
    +
    \alpha_\mathrm{max}\frac{\hbar\Omega}{4}(\hat{h}^\dagger+\hat{h})^2
    \nonumber
    \\
    & +U_h(\hat{x}_h,\hat{x}_g, t)
    + U_c(\hat{x}_c,\hat{x}_g, t)\,,
    \label{eqn:Hamiltonian_Second_Quantization}
\end{align}
where $\Omega = \sqrt{k / m}$ and $\alpha(t) = \kappa(t) / m\Omega^2 = \kappa(t) / k$.
Because the frequency of the hot mode is given by $\Omega_h = \sqrt{(k+\kappa) / m} = \Omega\sqrt{1+\alpha_\mathrm{max}}$, one gets $\alpha_\mathrm{max} = \omega^2 - 1$ with $\omega = \Omega_h / \Omega$ giving the ratio of compressed and uncompressed frequencies.
It is also convenient to express all the energies in terms of $\hbar\Omega$, time as $t = 2\pi \tau / \Omega$, and lengths in terms of the corresponding quantum oscillator length to get

\begin{align}
    \hat{H}(\tau) &=\sum_{d = c,h,g} \left(\hat{d}^\dagger \hat{d} + \frac{1}{2}\right)
    \nonumber
    \\
    &+
    \frac{\alpha(\tau)}{4}(\hat{g}^\dagger+\hat{g})^2
    +
    \frac{\alpha_\mathrm{max}}{4}(\hat{h}^\dagger+\hat{h})^2
    \nonumber
    \\
    & +\Phi_h(\hat{x}_h,\hat{x}_g, \tau)
    + \Phi_c(\hat{x}_c,\hat{x}_g, \tau)\,,
    \label{eqn:Hamiltonian_Unitless}
\end{align}
where $\Phi = U / \hbar\Omega$.

It is worth noting that only some terms must be included from Eq.~\eqref{eqn:Hamiltonian_Unitless} during the engine operation.
For example, during the compression and expansion phases, the interaction terms are switched off as $\alpha(\tau)$ changes in time.
On the other hand, only one of the interaction terms is nonzero during the bath coupling phases, and the uncoupled bath can be ignored as it evolves independently.

\section{Compression and expansion}
\label{sec:Compression_and_expansion}

When the gas undergoes compression or expansion, it is disconnected from the baths and the relevant normal-ordered portion of Eq.~\eqref{eqn:Hamiltonian_Unitless} is

\begin{equation}
    \hat{H}(\tau) = \left(\hat{g}^\dagger \hat{g} + \frac{1}{2}\right)\left[1+\frac{\alpha(\tau)}{2}\right]
    +
    \frac{\alpha(\tau)}{4}(\hat{g}^\dagger\hat{g}^\dagger+\hat{g}\hat{g})
    \label{eqn:Hamiltonian_Compression_Expansion}\,.
\end{equation}
The process is guided by the time evolution operator, given by the solution to the time-dependent Schr\"{o}dinger equation

\begin{equation}
    \frac{d}{d\tau}\hat{\mathcal{U}}(\tau, \tau') = -2\pi i \hat{H}(\tau) \hat{\mathcal{U}}(\tau, \tau')\,,
    \label{eqn:U_diff_eq}
\end{equation}
where the factor of $2\pi$ arises from the definition of $\tau$.
In this work, Eq.~\eqref{eqn:U_diff_eq} is solved using the fifth order Runge-Kutta method with the numerical benchmarks provided in Appendix~\ref{sec:Benchmarking}.

\begin{figure}
    \centering
    \includegraphics[width = \columnwidth]{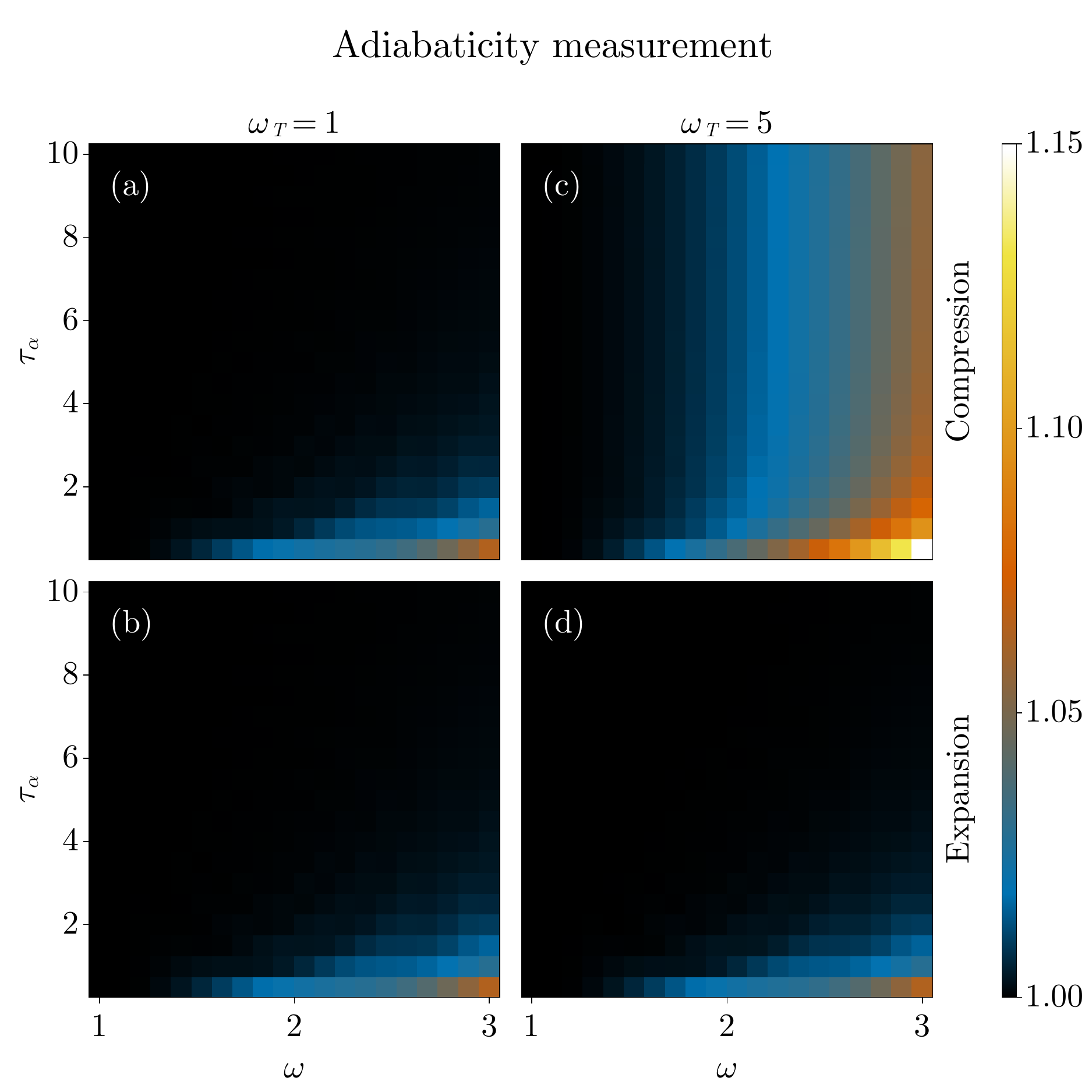}
    \caption{\emph{Nonadiabatic effects.} A system initialized at a 41-level thermal state at temperature $\omega_T = k_BT / \hbar\Omega$ with $\alpha = 0$ (top row) or $\alpha = \omega^2 -1$ (bottom row). Next, the state is either compressed (until it arrives at $\alpha = \omega^2-1$) or expanded (until $\alpha = 0$) uniformly over time $\tau_\alpha$. The final energy of the system is then divided by what would be the energy of the system if the modification were adiabatic. The nonadiabatic effects are most prevalent when the rate of the confinement change is high (corresponding to large $\omega$ and small $\tau_\alpha$). High-temperature compression demonstrates the effects of the finite Fock space, as discussed in the main text.}
    \label{fig:Adiabaticity}
\end{figure}

Although it is common to treat the compression/expansion phases of the quantum Otto cycle as adiabatic, the finite duration of the strokes is likely to give rise to nonadiabatic effects.
To quantify the amount of nonadiabaticity, it is convenient to proceed as follows.
Let a system, described by a Hamiltonian $\hat{H}_i$ start in some state $\hat{\rho}_i$.
If the system is compressed or expanded adiabatically, the final energy will be $E^\mathrm{A}_f = \mathrm{tr}\left[\hat{H}_f \hat{\rho}_i\right]$, where $\hat{H}_f$ is the final form of the Hamiltonian.
For a non-adiabatic modification, $E_f = \mathrm{tr}\left[\hat{H}_f\hat{\mathcal{U}} \hat{\rho}_i\hat{\mathcal{U}}^\dagger\right]$.
The level of nonadiabaticity is obtained by taking the ratio $E_f / E^\mathrm{A}_f\geq 1$.
Figure~\ref{fig:Adiabaticity} illustrates the effects of nonadiabatic tuning of thermal states at two temperatures which, as expected, are greatest for higher rates of change, corresponding to small $\tau_\alpha$, and large $\omega$.

In addition to illustrating the consequences of $d\alpha(\tau)/d\tau \neq 0$, Fig.~\ref{fig:Adiabaticity} highlights another aspect of this stroke that needs to be kept in mind when performing numerical simulations.
One can see that the difference between the finite-time and adiabatic results for the high-temperature compression phase, shown in Fig.~\ref{fig:Adiabaticity}(c), is greater than for all other cases.
This is a spurious effect arising from the finite size of the Fock space used here.
For any modification of the oscillator frequency, there is a finite probability of the oscillator transitioning to higher energy levels, especially during a compression.
At low temperatures, only the lowest-energy states are occupied at $\tau = 0$ so fewer Fock states are needed to correctly represent the compression.
As the temperature is increased, more levels become occupied, requiring a larger Fock space to allow the required transitions.
All panels in Fig.~\ref{fig:Adiabaticity} use 41 states and, for $\omega_T = 5$, this is not sufficient.
Halving the number of states aggravates the problem specifically for panel (c) while keeping the other panels relatively intact.
When demonstrating the engine operation in Sec.~\ref{sec:Engine_operation}, the hot bath's temperature will be set to $\omega_T = 5$, while the cold one will be kept at $\omega_T = 1/10$.
The fact that panels (a) and (d) do not change when reducing the number states indicates that this basis size is sufficient to avoid the finite-basis effects for these temperatures with $\omega \leq 3$ and $\tau_\alpha \geq 1$.

\section{Bath coupling}
\label{sec:Bath_coupling}

The purpose of the bath modes is to either add energy to or remove it from the gas mode.
The state of the baths does not have to take a particular form as long as the energy flow occurs in the right direction.
It is, however, convenient to consider thermal baths as illustrative examples so that their unnormalized density operators are $\hat{\rho}_b = e^{-\hat{H}_b / \omega_T}$.

Coupling the gas mode to a thermal bath will not, generally, set the gas to a thermal state.
However, repeating the process multiple times by introducing new identical thermal baths is expected to bring the gas to the bath temperature eventually.
One can use this physical intuition to check that the energy exchange between the modes functions appropriately before focusing on the engine itself.

When the gas couples to a bath, $\alpha$ remains fixed, leading to

\begin{align}
    \hat{H}(\tau) &= \left(\hat{g}^\dagger \hat{g} + \frac{1}{2}\right)\left(1+\frac{\alpha}{2}\right)
    +
    \frac{\alpha}{4}(\hat{g}^\dagger\hat{g}^\dagger+\hat{g}\hat{g})
    \nonumber
    \\
    &+\left(\hat{b}^\dagger \hat{b} + \frac{1}{2}\right)\left(1+\frac{\alpha}{2}\right)
    +
    \frac{\alpha}{4}(\hat{b}^\dagger\hat{b}^\dagger+\hat{b}\hat{b})
    \nonumber
    \\
    & +\Phi(\hat{x}_b,\hat{x}_g, \tau)\,.
    \label{eqn:Hamiltonian_Bath}
\end{align}
Assuming that the coupling between the gas and the baths is switched on and off quickly, it is reasonable to suppress the time argument inside the interaction term so that the matrix elements in the Fock space become 

\begin{align}
    &  \langle u, v| \Phi(\hat{x}_b,\hat{x}_g)|j,k\rangle
    \nonumber
    \\
     =
     &
     \int dx_bdx_g
      \Phi(x_b,x_g)
     \Psi_{j}(x_b)\Psi_{k}(x_g)
      \Psi_{u}^*(x_b)\Psi_{v}^*(x_g)\,,
      \label{eqn:Phi_Matrix}
\end{align}
where $\Psi_{n}(x) = \langle x|n\rangle$ are the harmonic oscillator wavefunctions.

As discussed in the introduction, the coupling between the modes is controlled using a central potential that can be switched on and off.
Equation~\eqref{eqn:Phi_Matrix} shows that if the coupling term is even in $x_b$ and $x_g$ (as is the case for a central potential aligned with the origin), only the states with the same parity in each of the modes can couple.
One can introduce coupling between more modes by positioning the extremum of the central potential away from the symmetric $(0,0)$ point so that the interaction term takes a general form $\Phi(\mathbf{r} - \mathbf{r}_0)$.
For illustration, it is convenient to use a Gaussian interaction $\Phi_0 \exp\left(-|\mathbf{r} - \mathbf{r}_0|^2/2\sigma^2\right) = \Phi_0 e^{-(x_g - x_{g,0})^2 / 2\sigma^2}e^{-(x_b - x_{b,0})^2 / 2\sigma^2}$
The advantages of this interaction are twofold.
First, its amplitude and extent are easily tunable.
Second, because the term is separable in $x_g$ and $x_b$, the integrals in Eq.~\eqref{eqn:Phi_Matrix} can be easily computed numerically.

If the extremum of the central potential is positioned diagonally from the origin ($x_{g,0} = x_{b,0} = x_0$), the procedure of computing the interaction matrix becomes particularly simple.
First, one obtains the matrix $\Phi_\mathrm{single}$ with elements $\langle j |e^{-(x - x_0)^2/2\sigma^2}|k\rangle$ for all the Fock states in the single-oscillator basis.
From this, the full interaction matrix becomes $\Phi = \Phi_0 \Phi_\mathrm{single}\otimes \Phi_\mathrm{single}$.

A set of four simulations is performed to demonstrate the energy flow between oscillator modes.
To this end, four thermal oscillator states using 41 energy levels are generated defined by $(\alpha, \omega_T)$ with $\alpha \in \{0, 8\}$ and $\omega_T \in \{\omega_T^\mathrm{cold} = 1, \omega_T^\mathrm{hot} = 5\}$.
The interaction width $\sigma$ and $x_0$ are set to 1.
The coupling strength will be allowed to vary, as described below.
For a given $\alpha$, one of the states is designated as the gas, while the other acts as the bath with the gas (bath) state denoted by $\hat{\rho}_g$ ($\hat{\rho}_b$).
The key idea is that by having the gas interact with a series of baths, the state of the gas should approach the bath state.
The trace distance is used to measure how close the gas is to $\hat{\rho}_b$.

For this demonstration, the gas will interact with twelve baths.
For the first five of them, the coupling strength $\Phi_0 = 1$, for the next four $\Phi_0 = 1/5$, and for the final three $\Phi_0 = 1/20$.
The reason for ramping down $\Phi_0$ is the reduction of energy stored in the interaction between the two oscillators: as the gas state approaches $\hat{\rho}_b$, most of the energy needs to be stored in the oscillator energy, not in the coupling term between the gas and the bath.

At $\tau = 0$, as the gas is brought into contact with the first bath, the trace distance is calculated.
At this moment, the full state of the system is $\hat{\rho}_\mathrm{total}^0 = \hat{\rho}_g \otimes \hat{\rho}_b$.
Because the Hamiltonian does not vary in time, $\hat{\mathcal{U}}(\delta\tau) = e^{-2\pi i \hat{H}\delta\tau}$ can be calculated exactly for $\Phi_0 = 1$ and $\delta\tau = 5$ so that the state of the composite system at later times is given by $\hat{\rho}_\mathrm{total}(n\times \delta \tau) = \hat{\mathcal{U}}^n \hat{\rho}^0_\mathrm{total}(\hat{\mathcal{U}}^\dagger)^n$.
For $1\leq n \leq 10$, the partial trace of the gas with respect to the bath $\mathrm{tr}_b[\hat{\rho}_\mathrm{total}(n\times \delta\tau)]$ is computed and its trace distance to $\hat{\rho}_b$ is calculated to show how the reduced density operator evolves while the gas is coupled to the bath.
At $n = 10$, the bath is traced out and the gas becomes coupled to a new bath in state $\hat{\rho}_b$ so that the total state is $\hat{\rho}_\mathrm{total}= \mathrm{tr}_b[\hat{\mathcal{U}}^{10}\hat{\rho}_g \otimes \hat{\rho}_b(\hat{\mathcal{U}}^\dagger)^{10}]\otimes \hat{\rho}_b$.
The process is repeated for each of the twelve bath using the $\hat{\mathcal{U}}$ for the appropriate $\Phi_0$.

\begin{figure}
    \centering
    \includegraphics[width = \columnwidth]{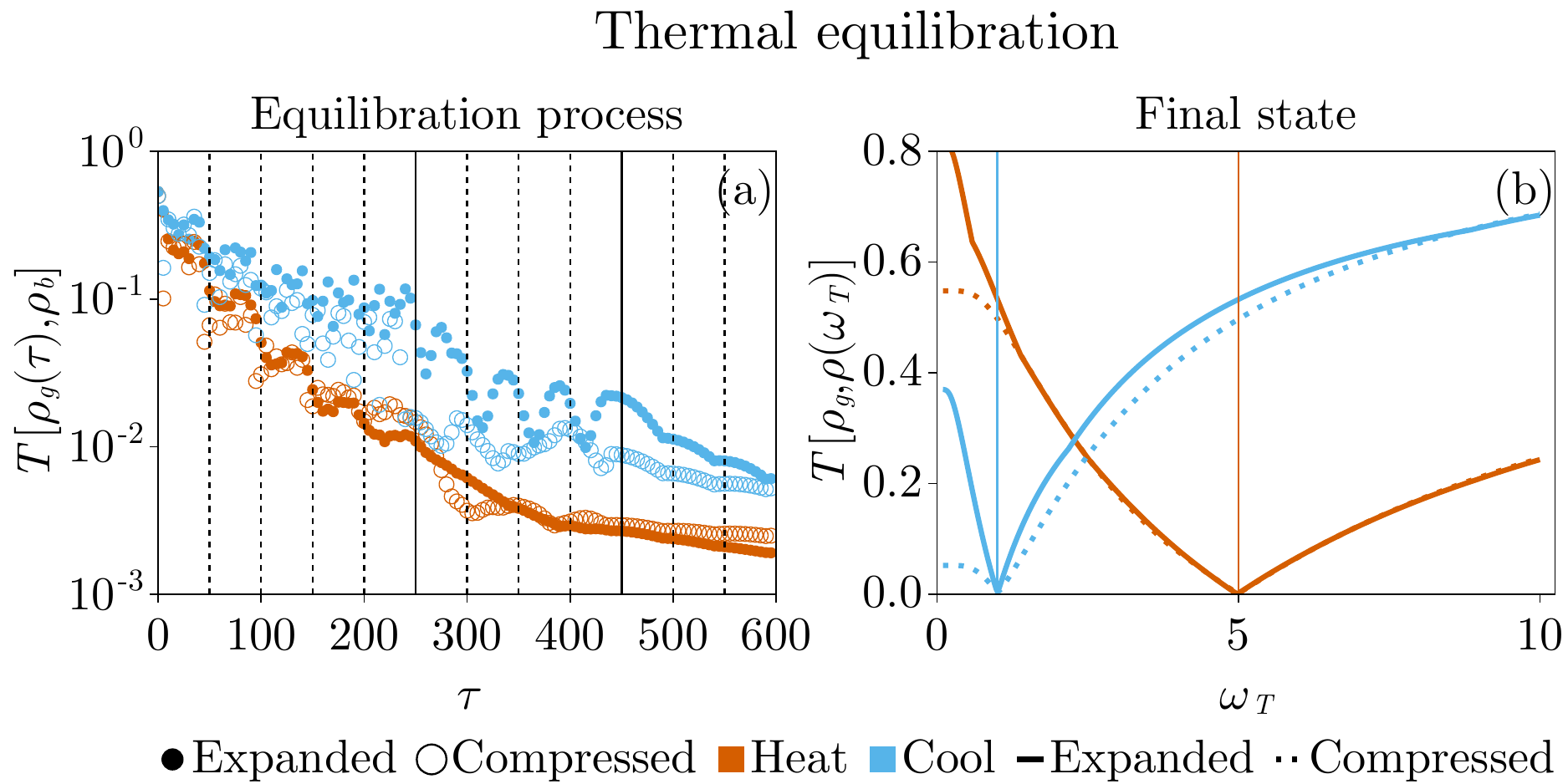}
    \caption{\emph{Interaction with a series of baths.} (a) Trace distance between the gas and bath modes as a function of time. The density operators include 41 Fock states. ``Expanded" (``compressed") corresponds to $\alpha = 0$ ($\alpha = 8$). ``Heat" means that the oscillator starts at $\omega_T = 1$ and the bath is at $\omega_T = 5$; for ``cool" the temperatures are switched. The interaction between the modes is Gaussian with $\sigma = 1$ and offset $x_0 = 1$. The interaction strength starts with $1$, later being reduced to $1/5$ and, finally, to $1/20$. The vertical lines mark the times when the bath is switched our for a new one. For dashed lines, $\Phi_0$ is unchanged; solid lines show where the interaction strength is reduced. (b) Trace distance between the final states from panel (a) and thermal states for a range of $\omega_T$. Vertical lines indicate the temperatures of the cold and hot baths. These lines coincide with the minima of the distance curves, indicating that the final gas state approaches the correct thermal state.}
    \label{fig:Equilibration}
\end{figure}

The computed trace distance as a function of time is plotted in Fig.~\ref{fig:Equilibration}(a).
One can see that the trace distance is less than $10^{-2}$ at the end of the simulation for all four configurations.
The final distance is smaller for the hot bath because, for a given $\Phi_0$, the interaction energy is proportionally smaller compared to the average oscillator energy at large temperatures.
One can reduce the distance for small $\omega_T$ by further lowering $\Phi_0$.

Figure~\ref{fig:Equilibration}(a) shows that the gas state approaches the bath.
However, it is not evident that the gas is not also approaching some temperature different the bath's $\omega_T$.
One can verify that the gas is indeed not drifting towards some wrong temperature by computing the trace distance between the final gas state and thermal states for a range of $\omega_T$ with the corresponding $\alpha$.
The results, shown in Fig.~\ref{fig:Equilibration}(b) confirm that the distance is smallest when $\omega_T$ equals the bath temperature.
Thus, even though the gas state is not quite thermal [because of the nonzero trace distance in panel (a)], the thermal state that it is closest to has the correct temperature.

It is important to note that the thermal equilibration protocol is somewhat ad hoc with the aim to demonstrate the general process, not make the energy transfer fast or efficient.
The contact times, $\sigma$, $x_0$, as well as the values of $\Phi_0$ were not optimized.
A more carefully designed procedure can speed up the process and bring the gas closer to $\hat{\rho}_b$.
The most important result of this section is that the coupling model used here results in the correct energy transfer between oscillator modes, validating its use in the engine description.

\section{Engine operation}
\label{sec:Engine_operation}

Having demonstrated the compression and expansion strokes of the engine, as well as the correct energy exchange with the baths, it is now possible to turn to the engine operation.
The main goal for this section to is confirm that the engine states are cyclical and that work can be extracted from the gas mode.

Without the loss of generality, it is convenient to have the cycle begin with the gas compression.
Thus, if the state of the gas in the beginning of the $n$th cycle is given by $\hat{\rho}_n$, the state of the gas in the beginning of the following cycle is

\begin{align}
    \hat{\rho}_{n+1} =\mathrm{tr}_b \left[\hat{\mathcal{C}} \left\{\hat{\mathcal{U}}^\dagger\mathrm{tr}_b\left[\hat{\mathcal{H}}\left(\hat{\mathcal{U}}\hat{\rho}_n\hat{\mathcal{U}}^\dagger \otimes \hat{\rho}_h\right)\hat{\mathcal{H}}^\dagger\right]\hat{\mathcal{U}}\otimes \hat{\rho}_c\right\}\hat{\mathcal{C}}^\dagger\right]\,.
    \label{eqn:Cycle}
\end{align}
Equation~\eqref{eqn:Cycle} should be read from the inside outward to follow the transformations that the gas undergoes.
First, sandwiching $\hat{\rho}_n$ between $\hat{\mathcal{U}}$ and its conjugate compresses the gas.
Next, the compressed gas is coupled to the hot bath in thermal state $\hat{\rho}_h$, as shown by the tensor product.
After that, the composite system is allowed to evolve in time by applying operators $\hat{\mathcal{H}}$ and $\hat{\mathcal{H}}^\dagger$, followed by a decoupling represented by the partial trace operator $\mathrm{tr}_b$.
Then, the gas is expanded, as can be seen by the reversed application of $\hat{\mathcal{U}}^\dagger$ and $\hat{\mathcal{U}}$, and connected to a cold bath $\hat{\rho}_c$.
Following the evolution guided by $\hat{\mathcal{C}}$, the gas is finally separated from the bath by the partial trace operator, completing the cycle.

For a particular engine configuration, $\hat{\mathcal{C}}$, $\hat{\mathcal{H}}$, $\hat{\mathcal{U}}$, and $\hat{\rho}_{c/h}$ need to be computed only once.
The first two operators are calculated by multiplying the total Hamiltonian by $-2\pi i \tau_\mathrm{contact}$ and exponentiating the result, where $\tau_\mathrm{contact}$ is the gas-bath interaction time.
$\hat{\mathcal{U}}$ is computed using the same Runge-Kutta scheme as was employed in Sec.~\ref{sec:Compression_and_expansion} for a linear compression.

\begin{figure}
    \centering
    \includegraphics[width = \columnwidth]{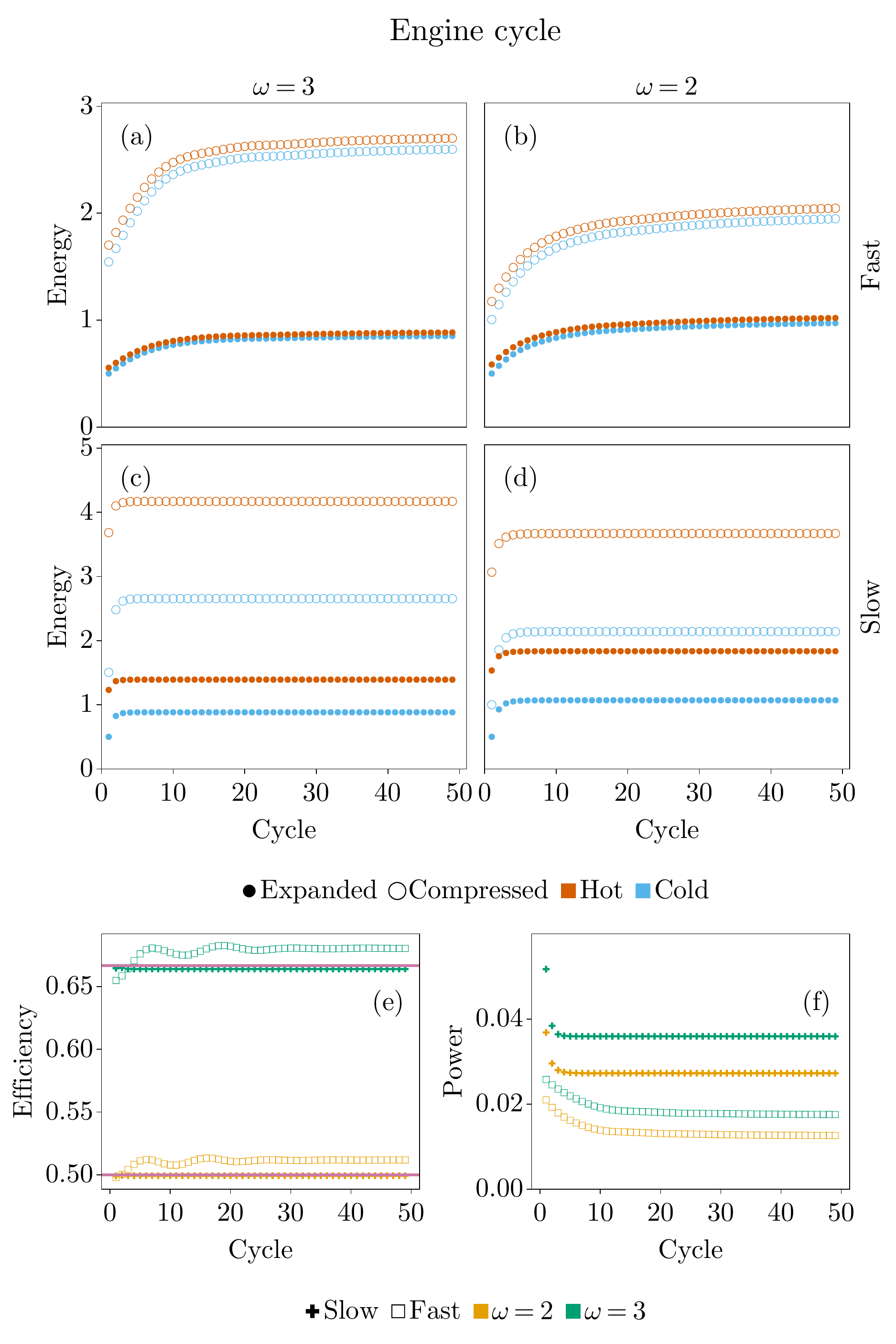}
    \caption{\emph{Engine operation}. (a)-(d) Energies of the gas in an Otto engine at each phase of the cycle as a function of time. For the fast engine (top row), the time of each stroke is 1. For the slow engine (bottom row), the time of expansion/compression is 4, while the bath contact time is 10. The system is evolved following Eq.~\eqref{eqn:Cycle} with $\omega_T^\mathrm{hot} = 5$ and $\omega_T^\mathrm{cold} = 1/10$. The interaction between the baths and the gas is the same as in Fig.~\ref{fig:Equilibration} with $\Phi_0 = 1$. For the compression and expansion phases, $\hat{\mathcal{U}}$ is calculated for a 41-state basis. (e) Efficiency for each of the engines as a function of the cycle number, computed by dividing the total work output by heat input. (f) The power of the engines, obtained by dividing the total work output by the duration of a cycle.}
    \label{fig:Operation}
\end{figure}

Four configurations are chosen for demonstration with two different $\omega$'s (2 and 3) and two different sets of times used for expansion/compression and bath contact.
For the first pair, the duration of all strokes of the cycle is equal to $1$.
For the second pair, the expansion/compression time is set to 4, while the bath contact time is 10.
Hence, there are two ``fast" and two ``slow" cycles.
The bath temperatures are the same for all realizations with $\omega_T^\mathrm{cold} = 1/10$ and $\omega_T^\mathrm{hot} = 5$ and the basis consists of 41 states.

For each engine, the gas is initialized in the thermal state at $\omega_T^\mathrm{cold}$.
Next, it is taken 50 times through the cycle described by Eq.~\eqref{eqn:Cycle}.
At the end of each stroke, the energy of the gas is calculated by taking the trace of the product of the gas density operator and the Hamiltonian in Eq.~\eqref{eqn:Hamiltonian_Compression_Expansion} with $\alpha = 0$ or $\alpha = \omega^2 -1$, as appropriate.

The computed energies for the four realizations are given in Fig.~\ref{fig:Operation}(a)-(d).
The $x$-coordinate labels the cycle and, for each cycle, the order of the points is ``expanded cold" $\rightarrow$ ``compressed cold" $\rightarrow$ ``compressed hot" $\rightarrow$ ``expanded hot," after which one moves to ``expanded cold" of the next cycle.

First, one can observe that, for the slow engine, the energies stabilize within a few cycles.
The fast engine, on the other hand, requires a substantially longer time.
Aside from this effect, the largest difference between the two speeds is the amount of energy transferred to and from the baths, as can be seen from the difference between ``compressed cold" $\rightarrow$ ``compressed hot" and ``expanded hot" $\rightarrow$ ``expanded cold" transitions.
Naturally, the $\omega = 3$ configuration demonstrates larger energy changes during the compression and expansion phases, as expected.

To quantify the efficiency of the engine, one first calculates the work output by the engine, given by $-\left[(E_\mathrm{exp\,hot} - E_\mathrm{comp\,hot}) + (E_\mathrm{comp\,cold} - E_\mathrm{exp\,cold})\right]$.
Dividing this value by the heat delivered by the hot bath $E_\mathrm{comp\,hot} - E_\mathrm{comp\,cold}$ yields the efficiency, plotted in Fig.~\ref{fig:Operation}(e).
The plot demonstrates that, for the slow engine, the efficiency is $\approx 1/2$ for $\omega = 2$ and $\approx 2/3$ for $\omega = 3$.
These values agree well with the expected efficiency of an Otto cycle that operates in a fully adiabatic regime with thermal baths, where the efficiency is given by $1 - \omega^{-1}$.~\citep{Bhattacharjee2021}
Curiously, the efficiency of the fast engine is slightly higher, but still close to these values.
It is worth noting that both engines operate close to the adiabatic regime, as can be seen from Fig.~\ref{fig:Adiabaticity} for the values of $\tau_\alpha$ employed.
For the heat exchange, on the other hand, even the slow engine is not expected to reach the temperature of the bath, as confirmed by Fig.~\ref{fig:Equilibration}.

As is clear from panels (a)-(d), the fast engine yields much less work per cycle.
However, given that its cycle is seven times slower, it is more appropriate to compare the power of the two setups, as is done in Fig.~\ref{fig:Operation}(f) by dividing the work output by the cycle period.
This figure shows that despite a higher efficiency, the fast engine delivers less power by about a factor of two.

The most important message of this section is that the engine reaches a stable cycle and its efficiency agrees well with the predicted value both in a slow and fast operation regimes.
Additionally, for the realizations here, the factor that reduces the work output of the engine is not the nonadiabatic effects associated with fast compressing/expansion, but a shorter contact with the bath, limiting the amount of energy transferred.

\section{Summary}
\label{sec:Summary}

This work has introduced and simulated a realization of a quantum Otto engine comprising of a single three-dimensional harmonic oscillator.
One of the modes of the oscillator functions as a compressible working fluid, while the others act as hot and cold reservoirs.
The coupling between the baths and the working fluid is controlled by a nonlinear external potential.
Individual finite-time strokes of the engine were simulated numerically to explore the role of adiabaticity during the compression and expansion phases, as well as the energy flow during thermal contact with the baths.
It has been shown that even for a ninefold increase of the working fluid's force constant, performing the compression and expansion over a few oscillator periods essentially eliminates the nonadiabatic effects.
Additionally, it has been confirmed that having the working fluid interact with a series of baths eventually brings the working fluid to the bath temperature, as expected.
Finally, the study demonstrates that the working fluid reaches a stable state as it goes through multiple engine cycles.
The efficiency obtained here agrees well with the theoretically predicted value for the quantum Otto cycle operating in the adiabatic regime with thermal reservoirs.

There are several research directions that this study opens up.
In this work, it was assumed that the baths manage to reach thermal states while they are decoupled from the working fluid.
It is worth investigating the importance of the state being thermal and how the operation of the engine changes if it is not.
More importantly, because the bath modes become ``reset" while decoupled, they act as ancillae to connect the working fluid to thermodynamic reservoirs.
If the bath modes do not have to be in thermal states and simply need to be able to exchange energy with the working fluid, it is interesting to explore the possibility of the confining potentials acting as the energy reservoirs.
For example, the cold mode could be laser cooled to remove excess energy, while the hot mode experiences an external driving force that contributes energy to it.
As an extension, it is useful to explore the possibility of keeping the bath modes connected to their respective reservoirs throughout the engine operation.

Finally, the operation protocol presented here was not optimized for power or efficiency.
It would be useful to determine which parameters enhance the energy and power output of the engine.

\acknowledgements

The author acknowledges the National Research Foundation, Prime Minister Office, Singapore, under its Medium Sized Centre Programme and the support by Yale-NUS College (through Start-up Grant).
The author is grateful to Keian Noori and Silvia Lara for their input and discussion.

\appendix

\section{Numerical benchmarking}
\label{sec:Benchmarking}

\begin{figure}
    \centering
    \includegraphics[width = \columnwidth]{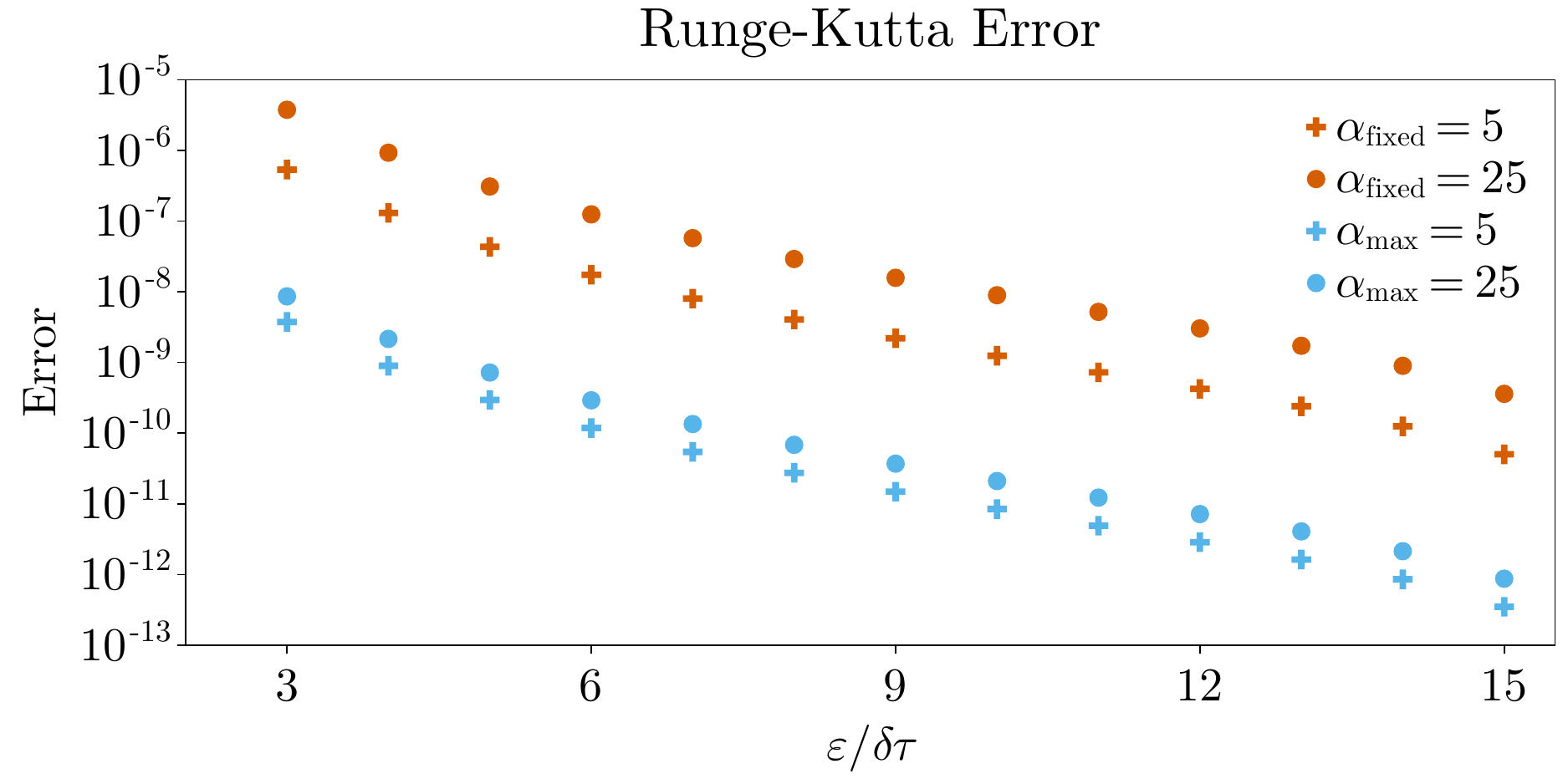}
    \caption{\emph{Numerical benchmarks.} A system is initialized in a 101-dimensional thermal mixed state with $\omega_T = 5$ and $\alpha = 0$ at $\tau = 0$. It is then evolved numerically until $\tau = 5$ using the fifth order Runge-Kutta method. For one set of simulations, $\alpha$ is set to a finite value at $\tau = 0^+$ and kept constant (labeled $\alpha_\mathrm{fixed}$ in the legend). For the second set, $\alpha$ increases linearly until it reaches its maximum value at $\tau = 5$ (labeled as $\alpha_\mathrm{max}$).
    The final state is calculated for $3\leq\varepsilon / \delta \tau\leq 16$ and the trace distance is obtained between the smallest time step and all others for each configuration. The trace distance is then taken as the error.}
    \label{fig:Error}
\end{figure}

To ensure numerical stability and physical realism, there are several guidelines that need to be followed when solving Eq.~\eqref{eqn:U_diff_eq}.
Most obviously, the Fock space has to be represented by a finite number of states.
When working with thermal states, the maximum level should be chosen so that $n_\mathrm{max}\gg \omega_T$ to guarantee the correct level occupancy, where $\omega_T = k_BT / \hbar\Omega$ is the thermal frequency.
With $n_\mathrm{max}$ fixed, the largest elements in the Hamiltonian are $\approx [1 + \alpha(\tau) / 2] n_\mathrm{max}$.
Consequently, the time step $\delta\tau$ has to be substantially smaller than the period associated with this frequency: $2\pi [1 + \alpha(\tau) / 2] n_\mathrm{max} \ll 1/\delta\tau$.
Hence, one needs to guarantee that $\omega_T \ll n_\mathrm{max} \ll \left[2\pi \delta\tau(1+\alpha_\mathrm{max} / 2)\right]^{-1}$.
It is useful to define $\varepsilon = \left[2\pi n_\mathrm{max}(1 + \alpha_\mathrm{max}/2)\right]^{-1}$ so the time step requirement becomes $\delta \tau \ll \varepsilon$.

As the first step, it is important to demonstrate the accuracy of the Runge-Kutta approach by performing two benchmark procedures.
For the first one, $\alpha(0)\neq 0$ is kept fixed and the system is initialized in a thermal mixed state with the density operator $\hat{\rho}_0 = e^{-\hat{H}_0 / \omega_T} / \mathrm{tr}\left[e^{-\hat{H}_0 / \omega_T}\right]$ for $\hat{H}_0 = \hat{g}^\dagger\hat{g} + 1/2$.
To illustrate the dependence of the numerical error on the step size, Eq.~\eqref{eqn:U_diff_eq} is solved for several values of $\delta\tau$ and the final density operator is computed using these $\hat{\mathcal{U}}$'s.
By taking the trace distance between each of the results and the density operator obtained for the smallest $\delta\tau$, the convergence of the numerical results is observed.
One should note that, because the Hamiltonian does not change in time, the analytical form of $\hat{\mathcal{U}}(\tau, 0) = e^{-2\pi i\tau\hat{H}}$ so that $\hat{\rho}(\tau) = e^{-2\pi i\tau\hat{H}}\hat{\rho}_0e^{2\pi i\tau\hat{H}}$.
The reason for not comparing the numerical results to the known analytical form has to do with the second benchmarking procedure where $\alpha(\tau)$ starts at zero and increases to some maximum value.
In this case, the analytical result is generally not known and the best one can do is show the convergence of the numerical calculations.
Hence, the same procedure is employed even in the constant-$\alpha$ case for the sake of consistency.
The results for these two checks are given in Fig.~\ref{fig:Error}, showing that one can achieve errors less than one part per million without making the time step drastically smaller than $\varepsilon$.
As a balance between accuracy and speed, the time step for RK solutions in the main text is taken $\delta\tau = \varepsilon / 5$.


%

\end{document}